\documentclass[12pt]{article} 
\usepackage{graphicx,floatflt,amssymb,rotate}
\usepackage{mathrsfs} 

\begin{document}

\vskip 30pt  
 
\begin{center}  
{\Large \bf KK-parity non-conservation in UED confronts LHC data
} \\
\vspace*{1cm}  
\renewcommand{\thefootnote}{\fnsymbol{footnote}}  
{ 
%{\sf Anindya Datta\footnote{email: adphys@caluniv.ac.in}}, 
%{\sf Amitava Raychaudhuri\footnote{email: palitprof@gmail.com}},  
{\sf Avirup Shaw\footnote{email: avirup.cu@gmail.com}} 
} \\  
\vspace{10pt}  
{\small {\em Department of Physics, University of Calcutta,  
92 Acharya Prafulla Chandra Road,\\ Kolkata 700009, India}}\\ 
   
\normalsize

\end{center}

\begin{abstract}  
\noindent

Kaluza-Klein (KK) parity can be violated in five-dimensional universal extra dimensional model with boundary-localized (kinetic or mass) terms (BLTs) at the fixed points of $S^1/Z_2$ orbifold. In this framework we study the resonant production of Kaluza-Klein excitations of the neutral electroweak gauge bosons at the LHC and their decay into an electron-positron pair or a muon-antimuon pair. We use the results (first time in our knowledge) given by the LHC experiment to constrain the mass range of the first KK-excitation of the electroweak gauge bosons ($B^1\ \textrm{and} \ W_3^1$). {It is interesting to note that the LHC result puts an upper limit on the masses of the $n=1$ KK-leptons for positive values of BLT parameters and depending upon the mass of $\ell^{+}\ell^{-}$ resonance.}\\ 

%such that they lie in a very narrow band close to the masses of $B^1\ \textrm{and} \ W_3^1$.

\vskip 5pt \noindent  
\texttt{PACS Nos:~11.10.Kk, 14.80.Rt, 13.85.-t  } \\  
\texttt{Key Words:~~Universal Extra Dimension, Kaluza-Klein, LHC}  
\end{abstract}

\renewcommand{\thesection}{\Roman{section}}  
\setcounter{footnote}{0}  
\renewcommand{\thefootnote}{\arabic{footnote}}  

\section{Introduction}

Discovery of the Higgs boson at the Large Hadron Collider (LHC) at CERN is a milestone in the success of Standard Model (SM). However, there are still many unanswered questions and unsolved puzzles, ranging from dark matter to the hierarchy problem to the strong-CP problem. But there is no experimental result that can explain such unsolved problems with standard particle physics. Out of various interesting alternatives, supersymmetry (SUSY) and extra dimensional models are the most popular frameworks for going beyond the SM of particle physics.
In this work we consider a typical extra-dimensional model where all SM particles can access an extra space-like dimension $y$. We use the results \cite{atlas8T}  presented by the ATLAS Collaborations of search for a high-mass resonances decaying into the $\ell^{+}\ell^{-}$ ($\ell \equiv e \ \textrm{or} \ \mu$) pair to constrain the parameter space of such a model  where the lowest ($n = 1$) KK-excitations are unstable due to lack of any specified symmetry.

We are interested in a specific framework, called the Universal Extra Dimension (UED) \cite{acd} scenario, characterised by a single flat  extra space like dimension $y$ which is compactified on a circle $S^1$ of radius $R$ and has an imposed $Z_2$ symmetry ($y \rightarrow-y$) to accommodate chiral fermions, hence the compactified space is called $S^1/Z_2$ orbifold. From a four-dimensional viewpoint, every field will then have an infinite tower of KK-modes, the zero modes being identified as the SM states. In this orbifold a $\pi R$ amount of translation in $y$ direction leads to a conserved KK-parity given by $(-1)^n$. The conservation of KK-parity ensures that the lightest
$n = 1$ KK-particle called Lightest Kaluza-Klein (LKP) is absolutely stable and hence is a potential dark
matter candidate. As the masses of the SM particles being
small compared to $1/R$, hence this scenario leads to an almost degenerate particle spectrum at each KK-level. This mass degeneracy could be lifted by radiative corrections. Being an extra dimensional theory and hence being non-renormalizable, this can only be an effective theory characterised by a cut-off scale $\Lambda$. So at the two fixed points ($y = 0$ and $y = \pi R$) of  $S^1/Z_2$ orbifold, one can include
four-dimensional kinetic and/or mass terms for the KK-states. These terms are also required as counterterms for 
cut-off dependent loop-induced contributions \cite{georgi}
 of the five-dimensional theory.  In the minimal
Universal Extra-Dimensional Models (mUED) these
terms are fixed by requiring that the five-dimensional loop
contributions \cite{cms1, cms2} are exactly cancelled at the cut-off scale
$\Lambda$ and the boundary values of the corrections, e.g.,
logarithmic mass corrections of KK-particles, can be taken to be
zero at the scale $\Lambda$. There are several existing literature \cite{acd}, \cite{nath}-\cite{flacke2} in which we can find how the experimental results constrain the values of the two basic parameters ($R$ and $\Lambda$) of mUED theory.

%In existing literature \cite{acd}, \cite{nath,db,chk,buras,desh,santa,appel-yee,ewued, precision,collued, LHCout,ILC,flacke2} we have found that the predictions of the theory have been compared with experimental data to set bounds on two basic parameters ($R$ and $\Lambda$) of the mUED theory.

In this work we generate non-conservation of KK-parity\footnote{This is equivalent to R-parity violation in supersymmetry.} by adding unequal boundary terms at the two fixed boundary points. Consequently $n = 1$ KK-states are no longer stable. Hence the single production of  $n = 1$ KK-states and its subsequent decay into $n = 0$ states would be possible via this non-conservation of KK-parity. We will utilise this KK-parity-non-conserving coupling of the $B^1 (W_3^1)$ to a pair of
SM fermions ($n =0$ states) \cite{ddrs}  to calculate the (resonance) production cross section of $B^1 (W_3^1)$ in $pp$ collisions at the LHC (8 TeV) and its subsequent decays to $e^+ e^-$/$\mu^+ \mu^-$, assuming $B^1\ \textrm{and} \ W_3^1$ to be the lightest KK-particles. Once $B^1\ \textrm{and} \ W_3^1$ are produced via KK-parity-non-conserving coupling, the KK-parity-conserving decaying mode being kinematically disallowed, thus the $B^1\ \textrm{and} \ W_3^1$ decay to a pair of
zero-mode fermions via the same KK-parity-non-conserving coupling. A search for high-mass resonances based on 8 TeV LHC $pp$ collision data collected by the ATLAS and CMS have been reported in \cite{atlas8T} and \cite{cms8T} respectively. Refs. \cite{atlas8T} and \cite{cms8T} present the expected and observed exclusion upper limits on cross section times branching ratio at 95\% C.L. for the combined dielectron and dimuon channels for resonance search. In  this article we have used the above results to constrain the masses of the $n = 1$
level KK-fermions and $B^1\ (\ W_3^1)$ of the model. In an earlier article \cite{drs} we have reported  the production of $n=1$ KK-excitation of gluon and its subsequent decay to $t\bar{t}$ pair at the LHC. Both the production and decay are governed by KK-parity-non-conserving interaction. Constraints have been also derived by comparing the $t\bar{t}$ cross section with LHC data from CMS \cite{tt_cms8T} and ATLAS \cite{tt_atlas8T} Collaborations. 

%We would like to
%emphasize that this is the first effort to restrict the
%parameters in the electroweak sector of KK-parity-violating UED using existing LHC data.

The plan of this article is as follows. At first we present the relevant couplings and masses in the framework of UED with asymmetric boundary localized kinetic terms. We then review the expected $\ell^{+}\ell^{-}$ signal from the combined production of the $B^1\ \textrm{and} \ W_3^1$ at the LHC and their subsequent decay.
This is compared with the %CMS \cite{cms8T} and%
 ATLAS
\cite{atlas8T} 8 TeV results  and the
restrictions on the couplings and KK-excitation masses are exhibited.
Finally we will summarise the results in section V.

\section{KK-parity-non-conserving UED in a nutshell}

In non-minimal version of five-dimensional UED theory, we put boundary-localized
kinetic terms (BLKTs) \cite{ddrs}, \cite{Dvali} - 
%carena, delAguila,delAguila_STU, flacke, 
\cite{asesh} at the orbifold fixed points ($y=0$ and $y=\pi R$). If $\Psi_{L,R}$ are the free fermion fields, zero modes of which are the chiral projections of the SM fermions. In presence of BLKTs, the five-dimensional action can be written as \cite{ddrs}, \cite{schwinn}: 
 
\begin{eqnarray} 
S & = \int d^4x ~dy \left[ \bar{\Psi}_L i \Gamma^M \partial_M \Psi_L 
+ r^a_f \delta(y) {\phi} ^\dagger _L i \bar \sigma^\mu \partial_\mu \phi_L 
+ r^b_f \delta(y - \pi R) {\phi} ^\dagger _L i \bar \sigma^\mu
\partial_\mu \phi_L
\right. \nonumber \\
&  \left. + \bar {\Psi} _R i \Gamma^M \partial_M \Psi_R
+ r^a_f \delta(y) {\chi} ^\dagger _R i {\sigma}^\mu \partial_\mu \chi_R 
+ r^b_f \delta(y - \pi R) {\chi} ^\dagger _R i {\sigma}^\mu
\partial_\mu \chi_R
\right]  .
\label{faction}
\end{eqnarray} 
With $\sigma^\mu \equiv (I, \vec{\sigma})$ and $\bar{\sigma}^\mu
\equiv (I, -\vec{\sigma})$, $\vec{\sigma}$ being the $(2 \times
2)$ Pauli matrices. $r^a_f, r^b_f$ are the free BLKT parameters which are equal for $\Psi_L$
and $\Psi_R$ for the purpose of illustration.

The KK-decomposition of five-dimensional fermion fields using two component chiral spinors are introduced as\footnote{We use the chiral representation with $\gamma_5 = diag(-I, I)$.} \cite{ddrs}, \cite{schwinn}:

\begin{equation} 
\Psi_L(x,y) = \pmatrix{\phi_L(x,y) \cr \chi_L(x,y)} 
=   \sum^{\infty}_{n=0} \pmatrix{\phi_n(x) f_L^{n}(y) \cr \chi_n(x) g_L^{n}(y)}
\;\; , 
\label{fiveDL}
\end{equation} 
\begin{equation} 
\Psi_R(x,y) = \pmatrix{\phi_R(x,y) \cr \chi_R(x,y)} 
=   \sum^{\infty}_{n=0}  \pmatrix{\phi_n(x) f_R^{n}(y) \cr \chi_n(x) g_R^{n}(y)} 
\;\;  . 
\label{fiveDR}
\end{equation}

%Below we discuss the case of $\Psi_L$ in detail. The formulas 
%for $\Psi_R$ will be similar and can be obtained by
%making appropriate changes.

%Using Eq. (\ref{fiveDL}), Variation of the above action Eq. (\ref{faction}) leads to coupled equations for the $y$-dependent wave-functions \footnote{More details in the same notations and
%conventions can be found in \cite{ddrs}.}, $f_L^{n}, g_L^{n}$. 
%\begin{equation}
%\left[1 + r^a_f \delta(y) + r^b_f \delta(y - \pi R) \right] m_n f_L^n - 
%\partial_y g_L^n = 0,\;\;
%m_n g_L^n + \partial_y f_L^n = 0, \; (n = 0,1,2, \ldots).
%\end{equation}
%Eliminating $g_L^n$ one obtains the equations:
%\begin{eqnarray}
%\partial_y^2 f_L^n &+& \left[1 + r^a_f \delta(y) + r^b_f \delta(y - \pi R) 
%\right] m_n^2 f_L^n = 0 .
%\end{eqnarray}
%Now we drop the subscript $L$ on the wave-functions  and
%denote them simply by $f$ and $g$.

%The boundary conditions are \cite{carena}
%\begin{equation}
%f^n(y)|_{0^-} = f^n(y)|_{0^+},\;\; f^n(y)|_{\pi R^+} = f^n(y)|_{\pi R^-} , 
%\end{equation}
%\begin{equation}
%\frac{df^n}{dy}\bigg|_{0^+} - \frac{df^n}{dy}\bigg|_{0^-} = -r_f^a
%m_n^2 f^n(y)|_{0}, \;\;
%\frac{df^n}{dy}\bigg|_{\pi R^+} - \frac{df^n}{dy}\bigg|_{\pi R^-} = -r_f^b
%m_n^2 f^n(y)|_{\pi R} .
%\end{equation}

Using appropriate boundary conditions \cite{ddrs}, we can have the solutions for $f_L^n$ and $g_R^n$ which are simply denoted by $f$ and $g$ for illustrative purposes.:
\begin{eqnarray}
f^n(y) &=& N_n \left[ \cos (m_n y) - \frac{r_f^a m_n}{2} \sin (m_n
y) \right] \;,\;\;  0 \leq y < \pi R,   \nonumber \\ 
f^n(y) &=& N_n \left[ \cos (m_n y) + \frac{r_f^a m_n}{2} \sin (m_n
y) \right] \;,\;\; -\pi R \leq y < 0.
\label{sol1}
\end{eqnarray}
Where the KK-masses $m_n$ for  $n = 0,1, \ldots$ 
are solutions of the transcendental equation \cite{carena}:
\begin{equation} 
(r^a_f r^b_f ~m_n^2 - 4) \tan(m_n \pi R)= 2(r^a_f + r^b_f) m_n \;.
\label{trans1}
\end{equation}

The non-trivial wave-functions are combinations of a sine
and a cosine function which are different from case of mUED where they are either only
sine or cosine function. The departure of wave functions from mUED theory and the fact that the KK-masses are solutions of Eq. (\ref{trans1}) rather than just $n/R$ are the key features of this non-minimal Universal Extra Dimensional (nmUED) model.

In our analysis we study the KK-parity-non-conserving UED in two ways.
In the first case, we take equal strength of BLKTs (at two fixed point $y = 0$ and $y = \pi R$) for fermion, i.e., $r^a_f = r^b_f \equiv r_f$, while the other case has the BLKT at
one of the fixed points only:   $r^a_f \neq 0, ~r^b_f = 0$. In
the later situation Eq. (\ref{trans1}) becomes:
\begin{equation} 
\tan(m_n \pi R)=-\frac{r^a_f m_n}{2} .
\label{trans3f}
\end{equation}

%*************report*********************************

{In both cases the mass eigenvalues can be solved from the transcendental equations (Eqs. \ref{trans1} and \ref{trans3f}) using numerical technique.}

{For small values of $\frac{r^a_f}{R}$ $(<<1)$ the approximate KK-mass formula becomes (using Eq. \ref{trans3f}):}

\begin{equation}
{ 
m_n\approx\frac{n}{R}\left(\frac{1}{1+\frac{r^a_f}{2\pi R}}\right)\approx\frac{n}{R}\left(1-\frac{r^a_f}{2\pi R}\right).}
\label{apprx}
\end{equation}

{It is clear from the above expression that for $r^a_f>0$, the KK-mass diminishes with $r^a_f$. This result also holds good when the BLKTs are present at both the boundary points.}

$N_n$ being the normalisation constant and determined from
orthonormality condition \cite{ddrs}:

%The {\em orthonormal} solutions satisfy the condition:
\begin{equation}
\int dy \left[1 + r^a_f \delta(y) + r^b_f \delta(y - \pi R)
\right] ~f^n(y) ~f^m(y) = \delta^{n m},
%\int dy \left[1 + r_f \delta(y)\right] g_R^n g_R^m = \delta^{n m} .
\end{equation}

%When $r^a_f = r^b_f \equiv r_f$  
%\begin{equation}
% N_n = \sqrt{\frac{2}{\pi R}}\left[ \frac{1}{\sqrt{1 + \frac{r_f^2 m_n^2}{4} 
%+ \frac{r_f}{\pi R}}}\right].
%\label{norm1}
%\end{equation}
%For the other case when $r^b_f = 0$  we use $r^a_f \equiv r_f$
%and one has
%\begin{equation}
% N_n = \sqrt{\frac{2}{\pi R}}\left[ \frac{1}{\sqrt{1 + \frac{r_f^2 m_n^2}{4} 
%+ \frac{r_f}{2 \pi R}}}\right].
%\label{norm2}
%\end{equation}

and is given by:
\begin{equation}
 N_n = \sqrt{\frac{2}{\pi R}}\left[ \frac{1}{\sqrt{1 + \frac{r_f^2 m_n^2}{4} 
+ \frac{r_f}{\pi R}}}\right],
\label{norm1}
\end{equation}
for equal strength of boundary terms ($r^b_f = r^a_f\equiv r_f$).

And for the other situation when $r^b_f = 0$ and we use $r^a_f \equiv r_f$ one has:
\begin{equation}
 N_n = \sqrt{\frac{2}{\pi R}}\left[ \frac{1}{\sqrt{1 + \frac{r_f^2 m_n^2}{4} 
+ \frac{r_f}{2 \pi R}}}\right].
\label{norm2}
\end{equation}

Now it is evident from the Eqs. \ref{norm1} and \ref{norm2}  that, for  $\frac{r_f}{R}<-\pi$~(for double brane set up)~and $\frac{r_f}{R}<-2\pi$ (for single brane set up) the squared norm of zero mode solutions become negative. Moreover for $\frac{r_f}{R}=-\pi$~(for double brane set up)~and $\frac{r_f}{R}=-2\pi$ (for single brane set up) the solutions become divergent. Beyond these region the fields become ghost like consequently the values of $\frac{r_f}{R}$ beyond these should be avoided. However for simplicity we stick to positive values of BLKTs only in the rest of our analysis.

Our concern here only with the zero modes and the $n=1$
KK-wave-functions of the five-dimensional fermion fields.

%Now if we look at the five-dimensional electroweak gauge field $V_N ~(N = 0
%\ldots 4)$, the action for that filed with BLKT $r^a_V, r^b_V$ at the fixed points can be
%similarly written down, and following
%similar steps\footnote{These steps are discussed in detail in
%\cite{ddrs}.} that in the $V_4 = 0$ gauge, the electroweak gauge field has the
%KK-expansion:  
%\begin{equation} 
%V_{\mu}(x,y)=\sum^{\infty}_{n=0}V_{\mu}^{n}(x) a^n(y),
%\end{equation} 
%where the functions $a^n(y)$  are of the same form as Eq.
%(\ref{sol1}). In this case the five-dimensional contributions to
%the KK-electroweak gauge bosons mass\footnote{The KK modes also receive a contribution
%to their masses from spontaneous breaking of the electroweak
%symmetry, but we have not considered that contribution as they are negligible with respect to extra-dimensional contribution.}, $m_n$, satisfy
%\begin{equation}
%(r_V^a r_V^b m_{n}^{2}-4)~\tan \left(m_{n}\pi R\right) = 2(r_V^a+r_V^b)
%m_{n} \; .
%\label{trans2}
%\end{equation}
%which is identical to Eq. (\ref{trans1}) for fermions.

%When the BLKT present at one of the fixed point only, i.e., ($r^a_V \neq 0 , \;
%r^b_V = 0$) the transcendental equation (\ref{trans2}) reduces to 
%\begin{equation} 
%\tan(m_n \pi R)=-\frac{r^a_V m_n}{2} .
%\label{trans3}
%\end{equation} 
%This equation is the same as Eq. (\ref{trans3f}) for fermions
%with similar BLKT.

Masses and $y$- dependent wave functions for the electroweak gauge bosons are very similar to the fermions and can be obtained\footnote{The KK-modes of gauge bosons also receive a contribution
to their masses from spontaneous breaking of the electroweak
symmetry, but we have not considered that contribution as they are negligible with respect to extra-dimensional contribution.} in a similar manner. We do not repeat these in this article.
This can be readily available in \cite{ddrs}.

As KK-masses obtained from transcendental equations are similar for fermions and gauge bosons, so we use $r_\alpha^a, r_\alpha^b$ to pameterised the strengths of BLKT  with $\alpha = f$ (fermions) or $V$ (gauge bosons) for the purpose of discussions. It has been assumed in the following that the KK-quarks are either mass degenerate or heavier than the KK-leptons. However they do not enter in our analysis. For simplicity we have assumed that BLKTs for U(1) and SU(2) gauge bosons are same, so that $B^1$ and $W_3^1$ are degenerate in mass. We are only interested in the $n = 1$ state.

%%%%%%%%%%%%%%%%%%%%%%%%%%%%%%%%%%%%%%%%%%%%%%%%%%%%%%%%%%%%%%%%%%%
%\begin{figure}[thb] 
\begin{figure}[h] 
\begin{center} 
\includegraphics[scale=0.6, angle=0]{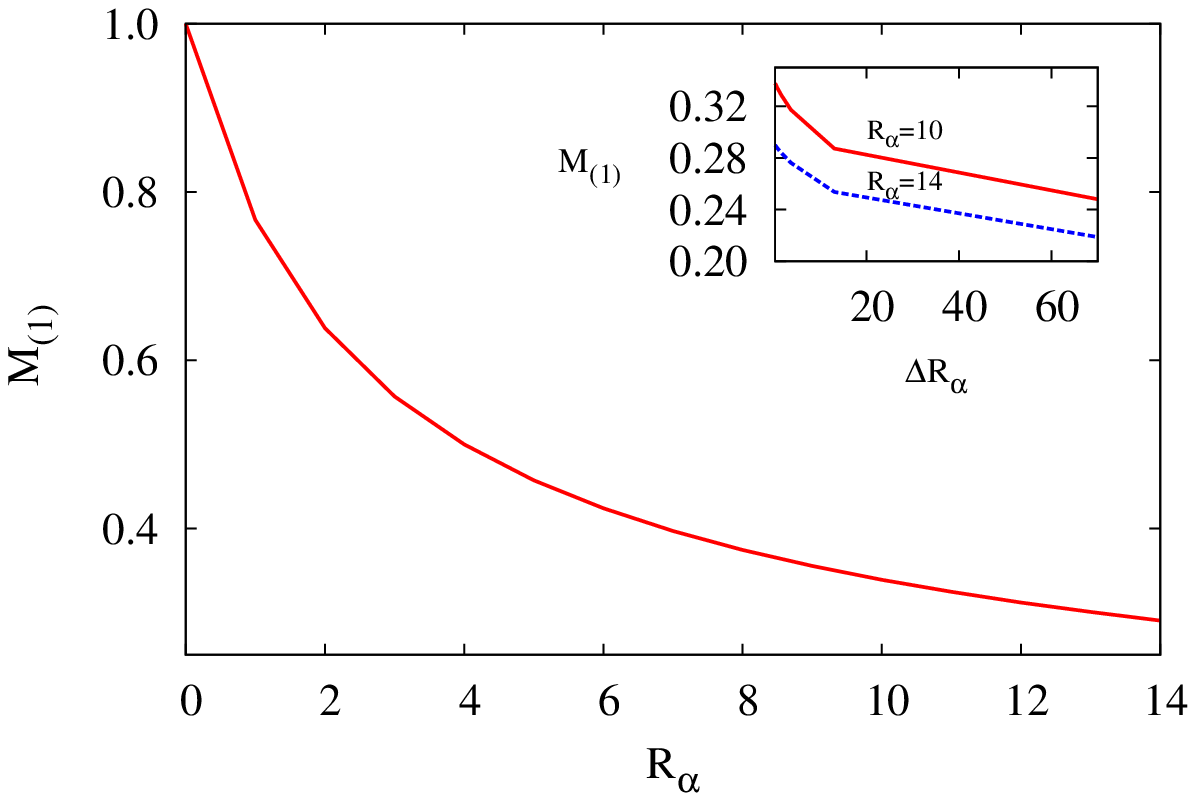}
\includegraphics[scale=.6, angle=0]{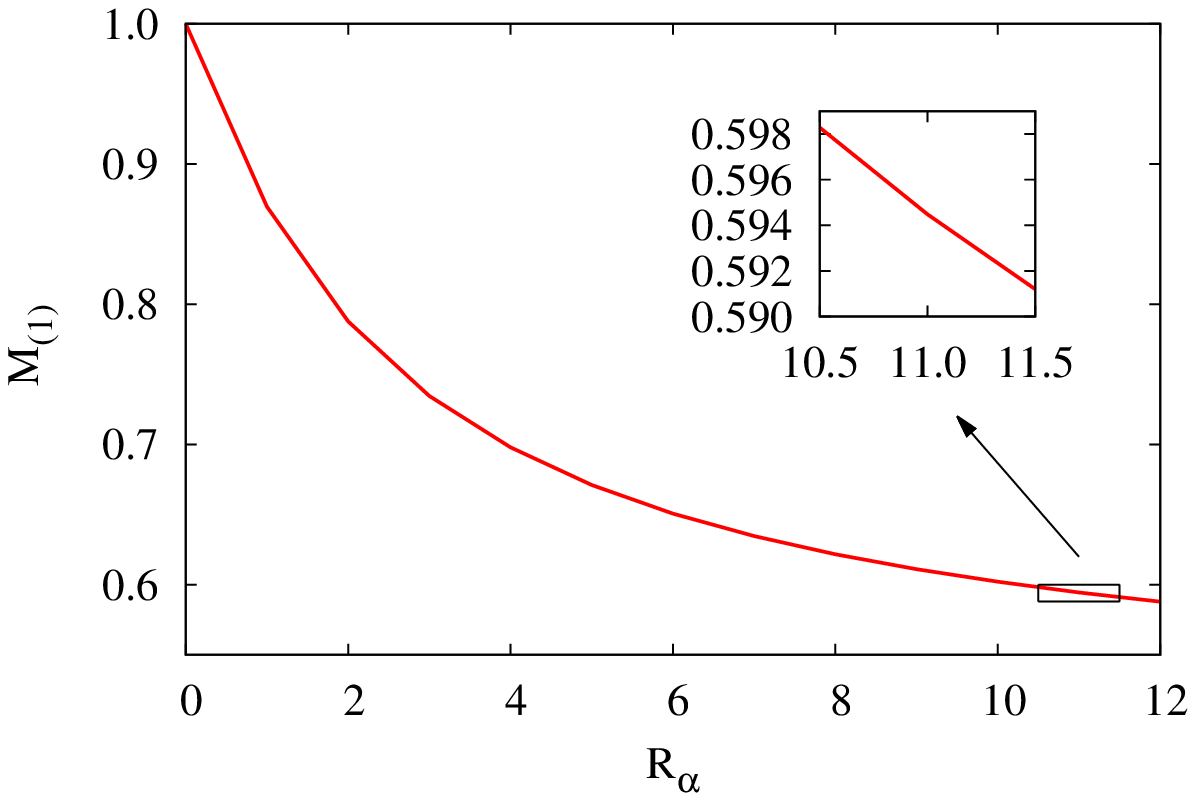}
\caption{Left panel: Dependence of $M_{(1)}\equiv m_{\alpha^{(1)}} R$ with respect to 
$R_\alpha \equiv r_\alpha^a/R$ when $r_\alpha^a =
r_\alpha^b$. In the inset is shown the variation of $M_{(1)}$ on
${\Delta R_\alpha} \equiv (r_\alpha^b - r_\alpha^a)/R$ for
two different $R_\alpha$. Right panel: $M_{(1)}$ with respect to
$R_\alpha$ when the BLKT is present only at the $y = 0$ fixed
point. The inset shows a magnified portion of the variation which gives the actual range of $R_\alpha$ that is
considered later. The insets of both panels shows small variation of $M_{(1)}$ with respect to BLKTs. Here $\alpha = f$
(fermions) and $V$ (gauge bosons).}
\label{KKmass} 
\end{center} 
\end{figure}

%----------------------

In Fig. \ref{KKmass} we have shown in plots a dimensionless quantity $M_{(1)}
\equiv m_{\alpha^{(1)}} R$, in the two different 
cases. The left panel reflects the mass profile for $n = 1$ KK-excitation when the BLKTs are presents at the two fixed points ($y=0$ and $y=\pi R$), while the right panel shows the case when the BLKTs are present only at $y = 0$. In both cases the KK-mass decreases  with the increasing values of BLKT parameter. The detailed illustrations of this non-trivial KK-mass dependence has been discussed in a previous article and can be found in \cite{drs}.

\section{Interacting coupling of \boldmath{$V^1$ ($B^1$ or $W_3^1$)} with zero-mode fermions} 

%We have now all the ingredients needed to calculate the
Coupling of the states $B^1$ or $W_3^1$ to two zero-mode fermions $f^0$ is given by, 
%Here $f^0$ could be SM quarks or leptons. It is given by
\begin{eqnarray} 
g_{V^{1}f^{0}f^{0}} &=&  
g_5(G)  ~\int^{\pi R}_{0} 
(1+r_{f}\{\delta(y)+\delta(y-\pi R)\})
f_{L}^{0}f_{L}^{0}a^{1} dy, \nonumber  \\
&=& g_5(G) ~\int^{\pi R}_{0} (1+r_{f}\{\delta(y)+\delta(y-\pi R)\})
g_{R}^{0}g_{R}^{0}a^{1} dy . 
\label{coup0}
\end{eqnarray}
Further, the five-dimensional gauge couplings $g_{5}(G)$ which appears above
is related to the usual coupling $g$ through 
\begin{equation}
g_{5}(G) = g ~\sqrt{\pi R \left(1+\frac{R^a_V+ R^b_V}{2\pi}\right)}.  
\end{equation}
%where
%\begin{equation}
%S_{G} = \left(1+\frac{R^a_V+ R^b_V}{2\pi}\right). 
%\end{equation}
We denote the zero-mode fermion wave-functions by $f_L^0$ and $g_R^0$ while the KK ($n = 1$)-gauge boson wave functions are denoted by $a^1$ depending on the values of chosen BLKTs.

%We remark in passing that, irrespective of the nature of the
%gauge bosons boundary terms, the coupling $g_{V^{1}V^{0}V^{0}}$ is always
%zero. Thus the resonant production of $V^1$ is initiated only by
%quarks and antiquarks in the colliding particles and the gluonic
%content of the proton plays no role. 

%\subsection{BLKT at both fixed points}
Let us first discuss the case in which BLKTs are presented at both the fixed points.
Here we assume for the fermions : $r_f^a = r_f^b = r_f$, 
but for the gauge bosons : $r_V^a \neq r_V^b$.
Using $y$-dependent wave-functions and proper normalisation \cite{ddrs} we get:

%of our interest here are found to be
%\begin{equation}
%f_{L}^{0} = g_{R}^{0} = \frac{1}{\sqrt{\pi R(1 + R_f/\pi)}}, 
%\end{equation}
%and
%\begin{equation}
%a^{1} = N_{V}^{1} 
%\left[\cos \left( \frac{M_{(1)}y}{R} \right)-\frac{R^{a}_V M_{(1)}}{2}
%\sin \left(\frac{M_{(1)}y}{R}\right)\right],
%\end{equation}
%with
%\begin{equation}
%N_{V}^{1} = \sqrt{\frac{1}{\pi R}}~\sqrt{\frac{8(4+M_{(1)}^{2}{R^b_V}^2)}
%{2\left(\frac{R^{a}_V+R^{b}_V}{\pi}\right)(4+M_{(1)}^{2}R^{a}_VR^{b}_V)
%+(4+M_{(1)}^{2}{R^{a}_V}^2)(4+M_{(1)}^{2} {R^{b}_V}^2)}}~~, 
%\end{equation}

\begin{eqnarray} 
g_{V^{1}f^{0}f^{0}} &=&  
 \frac{g_{5}(G)}{\left(1+\frac{R_{f}}{\pi}\right)}
\;N^1_G\;\left[\frac{\sin(\pi M_{(1)})}
{\pi M_{(1)}}\left\{1-\frac{M_{(1)}^{2}R^{a}_VR_{f}}{4}\right\}\right.
%\sqrt{\frac{8(4+M_{(1)}^{2}{R^b}^2)}
%{2\left(\frac{R^{a}+R^{b}}{\pi}\right)(4+M_{(1)}^{2}R^{a}R^{b})
%+(4+M_{(1)}^{2}{R^{a}}^2)(4+M_{(1)}^{2}{R^b}^2)}} 
\nonumber \\
&&
\left. +\frac{R^{a}_V}{2\pi}\left\{\cos(\pi M_{(1)})-1\right\}+
\frac{R_{f}}{2\pi}\left\{\cos( \pi M_{(1)})+1\right\}\right] ,
\label{coup1}
\end{eqnarray}
which vanishes\footnote{see Fig. 3 in \cite{ddrs}.} when $\Delta R_V = 0$.

Where $M_{(1)} \equiv m_{V^{(1)}} R$ is the scaled KK-mass, and
%\begin{equation}
$R_f \equiv r_f/R, \;\;  R^a_V \equiv r^a_V/R, \;\; {\rm and} \;\;
R^b_V \equiv r^b_V/R$ are the scaled dimensionless variables  defined earlier.
%\end{equation}
%Using the above  we get

Now we turn to the case which could be considered the
most asymmetric one, namely, the BLKT for the fermion and the gauge
boson are present only at the $y=0$ fixed point. We obtain for this case \cite{ddrs}:

%The $y$-dependent wave-functions in this case are
%\begin{equation} 
%f_L^{0} = g_R^{0} = \frac{1}{\sqrt{\pi R(1 + R_f/2 \pi)}}, 
%\end{equation}
%and
%\begin{equation}
%a^{1} = \sqrt{\frac{1}{\pi R}}~\sqrt\frac{2}{1+\left(\frac{R_V
%M_{(1)}}{2}\right)^2+\frac{R_V}{2\pi}}
%\left[\cos\left(\frac{M_{(1)}y}{R}\right)-
%\frac{R_V M_{(1)}}{2}\sin\left(\frac{M_{(1)} y}{R}\right) 
%\right] \;,
%\end{equation} 

\begin{eqnarray} 
g_{V^{1}f^{0}f^{0}} &=&  
\frac{\sqrt{2} ~g\sqrt{\left(1+\frac{R_{V}}{2\pi}\right)} }
{\left(1+\frac{R_{f}}{2\pi}\right) 
\sqrt{1+\left(\frac{R_V M_{(1)}}{2}\right)^2+\frac{R_V}{2\pi}}}  
\left(\frac{R_{f}-R_V}{2\pi}\right) \;,
\label{coup2}
\end{eqnarray} 

and this becomes zero if we put $R_V = R_f$. Here\footnote{As the BLKTs are present at only one fixed point so we use $R_f$ and $R_V$ with no superscript for fermions and gauge bosons respectively.} $R_V \equiv r_V/R$ and $R_f \equiv r_f/R$.

\begin{figure}
\begin{center}
\includegraphics[scale=1, angle=0]{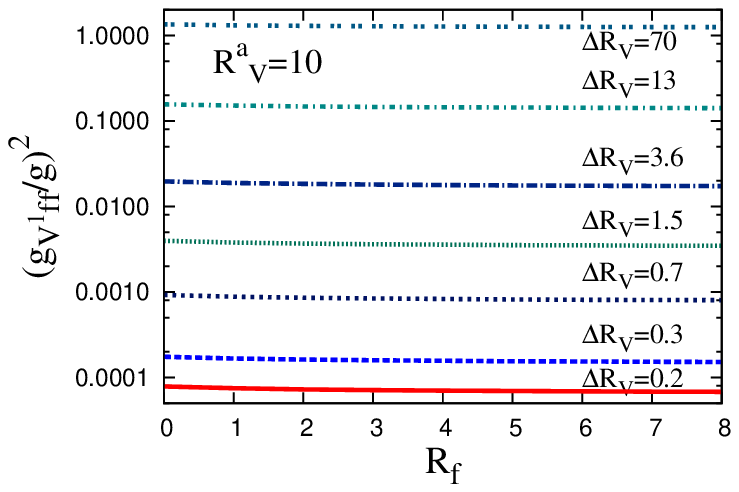}
\includegraphics[scale=1.0, angle=0]{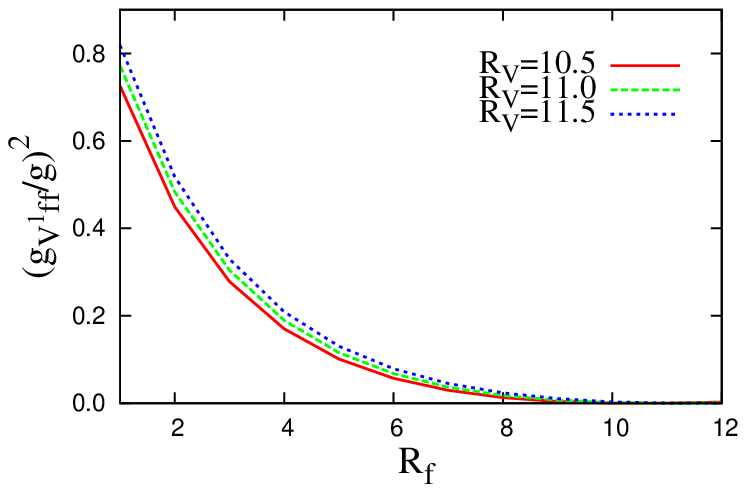}
\caption{Left panel: Dependence of the squared value of scaled KK-parity-non-conserving
coupling  between $B^1 (W_3^1)$ and a pair of zero-mode fermions with
$R_f \equiv R_f^a = R_f^b$ for different $\Delta R_V$, for  $R^a_V=10$.
Right panel: Dependence of the same coupling with
$R_f$ for different choices of $R_V$ when the fermion and gauge boson
BLKTs are present only at the $y = 0$ fixed point.}
\label{KKcoupling} 
\end{center} 
\end{figure}

The Fig. \ref{KKcoupling} depicts the KK-parity-non-conserving coupling strength in the two different cases. In the left panel (BLKTs are present at two fixed points $y=0$ and $y=\pi R$) we plot the square of the coupling for a fixed value\footnote{We have checked that the results are quite similar for the other
value of $R^a_V$ that we consider later.} of $R^a_V$ = 10 as a function of $R_f$ for several choices of $\Delta R_V$. The right panel shows the same thing with respect to $R_f$ (BLKTs are present only at the $y = 0$) for different values of $R_V$. In both cases the KK-parity-non-conserving coupling decreases  with the increasing values of fermion BLKT parameters. The detailed analysis of this coupling strength with respect to BLKT parameters can be found in a previous article \cite{drs}.

%for a fixed value\footnote{We have checked that
%the results are not significantly different for the other value
%of $R^a_V$ that we consider later.} of $R^a_V= 10$ 
%as a function of $R_f$ for several values of $\Delta
%R_V$. It is seen that the strength of the coupling decreases
%as $R_f$ increases while it increases with $\Delta R_V$. We emphasize that the KK-parity-non-conserving coupling gets smaller as $R_f$ tends towards $R_V$ i.e.,  as the splitting among the $n = 1$
%KK-excitations is decreased.  Also,  it can be argued if $R^a_V = R^b_V$, i.e., the gauge BLKTs are
%symmetric at $y = 0$ and $y = \pi R$  for  the gauge boson (as
%chosen for the fermions), the coupling in Eq. (\ref{coup1})
%vanishes. This can be traced to  a $y \longleftrightarrow (y-\pi
%R)$ $Z_2$-symmetry of the theory, which forbids an
%$n = 1$ state to couple exclusively to zero modes. In general, 
%$g_{V^{1}f^{0}f^{0}}$ decreases as $\Delta R_V$ gets smaller. 

%\subsection{BLKT at one fixed point}

%In the right panel  of Fig. \ref{KKcoupling} we plot the
%square of the coupling strength as a function of $R_f$ for
%several choices of $R_V$ for this alternative. In order to keep
%the $B^1 (W_3^1)$ lighter than the fermions of the same level
%we have kept $R_V > R_f$. It is to be noted that the strength of
%the coupling decreases as $R_f$ increases but it increases as
%$R_V$ increases. The coupling vanishes if $R_V=R_f$, as can be seen from Eq.
%(\ref{coup2}).

%----------------------

\section{Production and decay of \boldmath{$B^1 (W_3^1)$}  via KK-parity-non-conservation}

We are now in a position to discuss the main result of this paper. From now onwards for the SM particles we will not explicitly
write the KK-number $(n = 0)$ as a superscript. At the LHC we study the resonant
production of $B^1 (W_3^1)$, via the process $p p ~(q \bar q) \rightarrow B^1 (W_3^1)$ followed by $B^1 (W_3^1) \rightarrow \ell^{+}\ell^{-}$. This results into an $\ell^{+}\ell^{-}$ resonance at the $B^1 (W_3^1)$ mass.

The final state leads to two leptons $(e ~{\rm or} ~\mu)$, with
invariant mass peaked at $m_{V^{(1)}}$ which is the KK-mass ($n=1$) of gauge bosons.
% which is
%determined from Eqs.  (\ref{trans2}) and (\ref{trans3}) by $1/R$
%and the gauge boson BLKT  $R^{a,b}_{V}$. 
It should be noted that both
the production and the decay of $n=1$ KK-excitations of
electroweak gauge bosons are driven by KK-parity-non-conserving
couplings which depend on $R_f$, $R^a_V$  and $\Delta R_V$ ($R_f$, $R_V$ when BLKTs are present at only one fixed point). If in future such a signature is observed at the LHC, then it would be a good channel to measure such KK-parity-non-conserving couplings.

Both ATLAS \cite{atlas8T} and CMS \cite{cms8T} Collaborations have looked for a resonance decaying to $e^{+}e^{-}$/$\mu^{+}\mu^{-}$ pair in $pp$ collisions at 8 TeV in the LHC experiment. From the lack of observation of such a signal at 95\% C.L., upper bounds have been put on the cross section times  branching ratio of such a final state as a function of the mass of the resonance. The calculated
values of event rate in the KK-parity-non-conserving framework when compared to the experimental data set limits on the parameter space of the model. To get the most
up-to-date bounds we use the latest 8 TeV results\footnote{ATLAS results have been used in this paper as it used a higher accumulated data set. However we have checked that the limits derived from CMS \cite{cms8T} data are almost the same.} from ATLAS \cite{atlas8T}. 

% At this energy
%CMS has published \cite{cms8T} the analysis of 19.6$fb^{-1}$($e^{+}e^{-}$) and 20.6$fb^{-1}$($\mu^{+}\mu^{-}$)  of
%data while ATLAS has presented \cite{atlas8T} bounds from 20
%$fb^{-1}$ of data. We use the latter in our considerations\footnote{As the analysis presented by the ATLAS collaboration more convenient for our purpose.}. A specific upper bound on the event
%rate as quoted by  ATLAS \cite{atlas8T} therefore translates to
%constraints on the above parameters and thence to the mass of the
%KK-excitations of fermions.

Production of $B^1$ ($W_3^1$) (which we generically denote by $V^1$)
 in $pp$ collisions is driven $q
\bar q$ fusion. A compact form of the production cross section in proton proton collisions can be written as \cite{ddrs}:

\begin{equation}
\sigma (p p \rightarrow V^{1} + X) = \frac{4 \pi^2}{3 m_{V^{(1)}}^3}\;\sum_i 
\Gamma(V^{1} \rightarrow q_i \bar q_i)\;\int_\tau ^1 \frac{dx}{x}\;
\left[f_{\frac{q_{i}}{p}}(x,m_{V^{(1)}}^2) 
f_{\frac{{\bar q_{i}}}{p}}(\tau/x,m_{V^{(1)}}^2) + 
q_i \leftrightarrow \bar q_i \right].
\label{x-sections}
\end{equation}
Here, $q_i$ and $\bar{q_i}$ denote a generic quark and
the corresponding antiquark of the $i$-th flavour respectively.
$f_{\frac{q_{i}}{p}}$ ($f_{\frac{{\bar q_{i}}}{p}}$) is the parton distribution function for quark (antiquark) within a proton.
We define $\tau \equiv {m_{V^{(1)}}^2 / S_{PP}}$, where $\sqrt{S_{PP}}$
is the proton-proton centre of momentum energy. 
$\Gamma(V^{1} \rightarrow q_i \bar q_i)$ represents the decay
width of $V^1$ into the quark-antiquark pair and is given by 
%\begin{equation}
$\Gamma = \left[\frac{{g^{2}_{V^{1}q q}}}{32\pi}\right]\left[(Y^{q}_L)^2 + (Y^{q}_R)^2 \right]m_{B^{(1)}}$ (with $Y^q _L$ and $Y^q _R$ being the weak-hypercharges for the left- and right-chiral quarks) for $B^1$ and $\Gamma = \left[\frac{{g^{2}_{V^{1}q q}}}{32\pi}\right]m_{W_3^{(1)}}$ for the $W_3^1$. Here $g^{}_{V^1 q q}$ is the KK-parity-non-conserving coupling of the $V^1$ with the SM quarks -- see Eqs. (\ref{coup1}) and (\ref{coup2}). 

%Eq. (\ref{x-sections}) represents the lowest order result in QCD. We have not considered higher order contributions in our analysis
%and used it bearing in mind that QCD corrections usually enhance
%cross sections and so our results are probably conservative.

We use a parton-level
Monte Carlo code with parton distribution functions as
parametrised in CTEQ6L \cite{CTEQ} for determination of the 
numerical values of the cross sections. In our analysis we set the $pp$ centre of
momentum energy at 8 TeV and the factorisation scales (in the parton distributions)  at
$m_{V^{(1)}}$. To obtain the event rate one must multiply the cross sections with appropriate branching ratio\footnote{The branching ratio of $B^1\ (W_3^1)$ to $e^{+}e^{-}$ and $\mu^{+}\mu^{-}$ is approximately $\frac{30}{103}$ ($\frac{2}{21}$).} of $B^1$ or $W_3^1$ into $e^{+}e^{-}$/$\mu^{+}\mu^{-}$. {Here we have assumed without any loss of generality that $B^1$ and $W_3^1$ are lighter than the $n=1$ KK-excitation of the fermions.
Which implies that they are the lightest KK-particle and they can decay only to a pair of SM particles via KK-parity-non-conserving coupling--see Eqs. (\ref{coup1}) and (\ref{coup2}).}

At this end, let us comment about the values of the BLKT parameters used in our analysis. The BLKTs imposed in Eq. \ref{faction} are not five-dimensional operators in four-dimensional effective theory but 
some sort of boundary conditions on the respective fields at the orbifold fixed points. The masses (solutions of transcendental equations) 
and profiles in the $y$-directions for the fields are  consequences of these boundary conditions. In fact, four-dimensional effective theory only contains the {\em canonical }
kinetic terms for the fields and their KK-excitations along with their mutual interactions. The effect of BLKTs only shows up in
modifications of some of these couplings via an overlap integral ( see Eq. \ref{coup0}) and also in deviations of the masses from UED values of $n/R$ (in the $n$-th KK-level). So as long as these overlap integrals are not very large $( \lesssim 1 )$ we
do not have any problem with the convergence of perturbation series. In Fig. \ref{KKcoupling}, we have shown that the values of the overlap integrals (involves in the couplings determined by the five-dimensional wave functions) are $\lesssim 1$ for the entire range of the strengths of the BLKTs which have used in this article and never grow with these strengths.
  Furthermore it has been shown  in \cite{L_BLKTs} that theories with large strength of the BLKTs (relative to their natural cutoff scale, $\Lambda$) were found to be perturbatively consistent and are thus favoured. Such results in Ref. \cite{L_BLKTs} is in agreement with our observation that the numerical values of the overlap integral diminishes with increasing magnitude of BLKT coefficients ($r_{i}$'s).

We now present the main numerical results for two distinct cases, either BLKTs are present at both fixed points or only at one of the two, in following subsections.

\subsection{BLKTs are present at \boldmath{$y = 0$} and \boldmath{$y = \pi R$}}

\begin{figure}[h]  
\begin{center} 
\includegraphics[scale=1, angle=0]{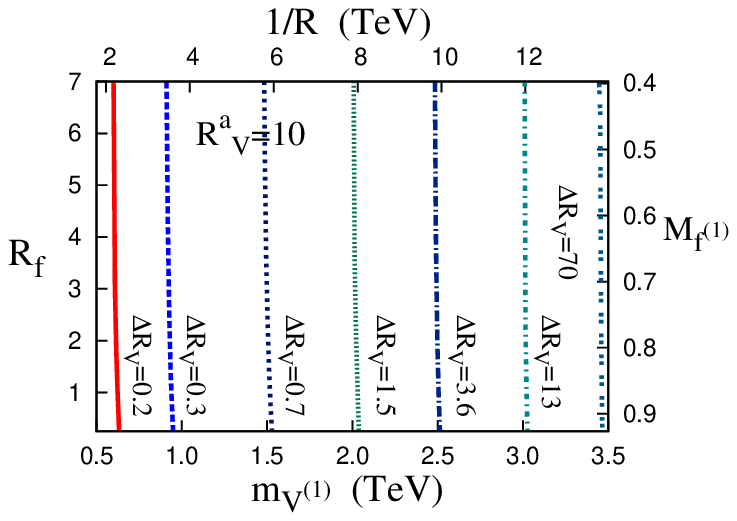}
\includegraphics[scale=1, angle=0]{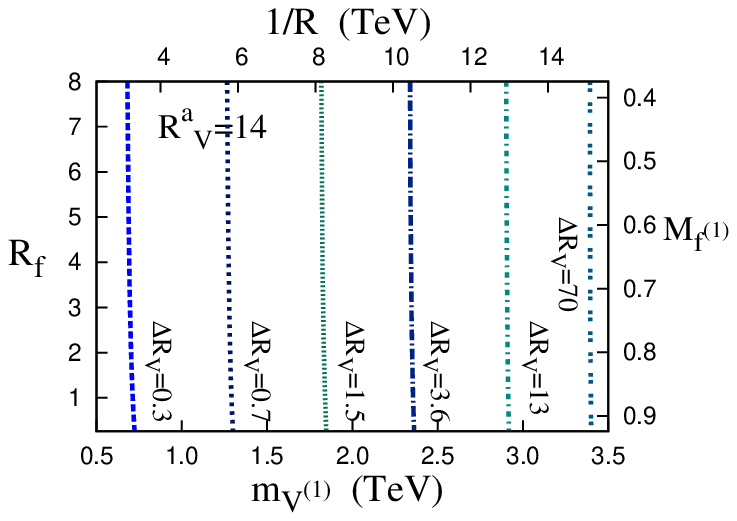}
%\includegraphics[scale=.36, angle=270]{rg_3_8.ps}
%\includegraphics[scale=.6, angle=270]{double_8TeV_CMS.ps}
%\vskip -20pt
\caption{95\% C.L. exclusion plots in the $m_{V^{(1)}} - R_f$ plane for several choices of $\Delta R_V$. Each panel corresponds to a specific value of
$R^a_V$. The region to the left of a given curve is excluded from the non-observation of a resonant $\ell^{+}\ell^{-}$ signal running at 8 TeV by ATLAS data \cite{atlas8T}. $1/R$ and $M_{f^{(1)}} =
m_{f^{(1)}} R$ are displayed in the upper and right-side axes
respectively (see text).}
\label{contours} 
\end{center} 
\end{figure}

In Fig. \ref{contours} we present the results for the case when the fermion BLKTs are symmetric at the two fixed points but unequal values of the gauge BLKTs break the KK-parity. Here we show the region of parameter space
excluded by the  ATLAS 8 TeV data \cite{atlas8T} for two different choices of $R_V^a$. Each panel depicts that the region to the left of a curve  in the
$m_{V^{(1)}}-R_f$ plane is excluded by the ATLAS data.

For a chosen $R^a_V$ there is an one-to-one correspondence of
$m_{V^{(1)}}$ with $1/R$ which is shown on the upper axis of the
panels, as the KK-mass is rather insensitive to $\Delta R_V$. Also, for any displayed value of $1/R$ we can estimate the first excitation of fermion KK-mass $M_{f^{(1)}} = m_{f^{(1)}} R$ ( plotted on right-side axis) which is determined by $R_f$. 

The exclusion plots can be understood easily in conjunction with Fig. \ref{KKmass} and Fig. \ref{KKcoupling}. For a given $\Delta R_V$ and $R^a_V$ the KK-parity-non-conserving couplings are almost insensitive to $R_f$. Thus $R_f$ has no steering on the production of $e^{+}e^{-}$/$\mu^{+}\mu^{-}$. The signal rate thus solely depend on $R^a_V$ and $\Delta R_V$. Coupling (and inturn signal strength) increases with $\Delta R_V$. Thus with higher and higher strength of KK-parity non-conservation one can exclude higher and higher masses (and higher values of $1/R$) as revealed in Fig. \ref{contours}.

%----------------------

%The implications of the above results on the $n = 1$ level
%KK mass spectrum can be extracted by considering them in
%conjunction with Figs. \ref{KKmass} and \ref{KKcoupling}. The
%limits on $m_{V^{(1)}}$ are essentially those on the $\ell^{+}\ell^{-}$
%resonance given in the data \cite{atlas8T}, and this puts an {\em upper} bound on the
%$n = 1$ fermion mass, which anyway has to be heavier than the $n = 1$ electroweak gauge boson in this model. Thus the fermion KK  excitation mass
%has to be in a limited range to agree with the LHC data.  This
%feature can be 
%illustrated by a few examples from Fig.
%\ref{contours}. From the left panel ($R_V^a = 10$) one
%finds that if $m_{V^{(1)}}$ = 0.60 TeV then $m_{f^{(1)}}$ is
%bounded from above by 1.64 TeV. If, on the other hand
%$m_{V^{(1)}}$ = 0.70 TeV the $n = 1$ fermion is constrained (see
%the right panel, $R_V^a = 14$) to lie between 0.70 and 2.70
%TeV. In Fig. \ref{contours} the two $R_V^a$ choices cover
%the entire ATLAS exclusion range of $\ell^{+}\ell^{-}$ resonance mass. As we have mentioned earlier that the mass of the $n = 1$ state for a particular $R_V$ always remains more than that corresponding to any larger $R_V$ for the entire variation of $\Delta R_V$.
%Therefore, any curve indicated by a particular $\Delta R_V$ moves towards the lower mass range for larger $R_V$, and this is quite evident from the two panels of Fig. \ref{contours}.

\subsection{BLKTs are present only at \boldmath{$y = 0$}}

\begin{figure}[h] 
\begin{center} 
\includegraphics[scale=1, angle=0]{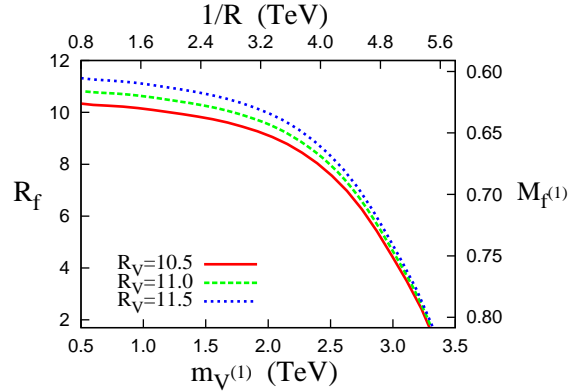}
\caption{95\% C.L. exclusion plots in the $m_{V^{(1)}} - R_f$ plane for
several choices of $R_V$. The region below a specific curve is
ruled out from the non-observation of a resonant $\ell^{+}\ell^{-}$ signal
in the 8 TeV run of LHC by 
%CMS \cite{cms8T} (left) and
 ATLAS \cite{atlas8T}. 
%(right)
$1/R$
and $M_{f^{(1)}} = m_{f^{(1)}} R$ are displayed in the upper and
right-side axes respectively (see text). 
%The $y$-axis ranges in the two panels are different.
}
\label{contoursS} 
\end{center} 
\end{figure}

Now let us concentrate on the case of fermion and gauge BLKTs at only
one fixed point. In this case we display the exclusion  curves in the $m_{V^{(1)}}-R_f$ plane for several choices of $R_V$ in Fig. \ref{contoursS}. The region below a curve has
been disfavoured by the ATLAS data. 
% As expected the CMS bounds
%based on 19.7 $fb^{-1}$ data are more restrictive than those from
%the 14.3 $fb^{-1}$ ATLAS result.

%%%%%%%%%%%%%%%%%%%%%%%%%%%%%%%%%%%%%%%%%%%%%%%%%%%%%%%%%%%%%%%%%%%
%\begin{figure}[thb] 

%----------------------

One can explain the Fig. \ref{contoursS} on the basis of Fig. \ref{KKmass} and Fig. \ref{KKcoupling}. It is revealed in the left panel of Fig. \ref{KKmass}, $M_{(1)}
\equiv m_{V^{(1)}} R$ has mild variation with
$R_V$. So we can take the mass of
$V^1$ to be approximately proportional to $1/R$ (the relevant values of $1/R$ are
displayed in the upper axis of the panel in Fig.
\ref{contoursS}). In our model we estimate the
cross section times branching ratio corresponding to any $m_{V^{(1)}}$, and comparing this with the ATLAS data we can have a specific value of $(R_V, R_f)$ pair on each curve via KK-parity-non-conserving coupling.  Alternatively, it is evident from the Fig. \ref{contoursS}, as $m_{V^{(1)}}$
increases, the production of the $B^1 (W_3^1)$ decreases. As a compensation, the KK-parity-non-conserving coupling must
increase as we increase the $m_{V^{(1)}}$. So it is clear from the right panel of
Fig. \ref{KKcoupling}, increasing value of KK-parity-non-conserving coupling is achieved by the higher values of $R_V$ for a fixed value of $R_f$. In this case also, the KK-fermion mass of first excitation can be obtained in a correlated way from the right-side axis of this plot.

{Let us pay some attention to  Fig. \ref{contoursS}. For a given curve (specified by a $R_V$), the allowed area in $m_{V^{(1)}} - R_f$ plane is bounded by the curve itself and a line parallel to $m_{V^{(1)}}$ axis corresponding to the value of $R_f$ determined by the specific value of $R_V$ of that curve.  The choice of the $R_f<R_V$ ensures mass hierarchy among KK-electroweak gauge boson and KK-leptons. So the bounded region implies that for a given value of $m_{V^{(1)}}$ , $R_f$ is bounded from below which in turn imposes an upper limit on the mass of the $n = 1$ KK-lepton. } 
%This is the case in Fig.
%\ref{contoursS}.

%Here we study above figure on the basis of some numerical values of masses of first excited KK-level. For example, if
%$m_{V^{(1)}}$ = 500 GeV then depending on whether $R_V$ is 10.5,
%11.0, or 11.5 the upper bound on the $n = 1$ fermion mass is 500.9, 492.7
%or 484.4 GeV. For a heavier $m_{V^{(1)}}$ of mass 2 TeV one
%finds that the upper bound on the corresponding fermion excitation
%is 2.15 TeV for $R_V = 10.5$, 2.12 TeV for $R_V = 11.0$ and 2.09 TeV for $R_V = 11.5$ respectively. These examples indicate that in this scenario, the $n = 1$ fermions and gauge bosons  have to be quasi-degenerate to tally with the LHC observations. With one assumption that at $n=1$ level  KK-fermions are heavier than $B^1$ and $W_3^1$, this result is remarkable in the sense that $n=1$ KK-fermion masses are thus constrained to be in a very narrow band.

%

\section{Conclusions}
In summary, we have investigated the phenomenology of KK-parity non-conservation in the UED model where all the SM  fields propagate in 4+1 dimensional space time. We have achieved this non-conservation due to inclusion of asymmetric\footnote{Symmetric BLTs leads to conserved $Z_2$ symmetry, hence $n = 1$ KK-particles is stable and can be a dark matter candidate \cite{dm,ddrs_dm}.} BLTs at the two fixed points of this orbifold. These boundary (kinetic in our case) terms can be thought of as a cut-off ($\Lambda$) dependent log divergent radiative corrections \cite{cms1} which remove the degeneracy in the KK-mass spectrum of the effective 3+1 dimensional theory.

%There are two distinct cases of choosing the four-dimensional kinetic terms at
%the fixed points. In both cases we end up with
%a theory where the spectrum of KK-particles and the couplings can
%be drastically different from mUED. In 
%the first, the BLKTs are of equal strength at both fixed points
%($y = 0$ and $y = \pi R$), which survives the $Z_2$ symmetry $y
%\longleftrightarrow (y - \pi R)$ , hence the lightest among the $n =
%1$ KK-particles is stable and can be a dark matter candidate \cite{dm,ddrs_dm}.
%The other alternative is to allow the BLKTs at the two fixed
%points to be of unequal strengths.  This  will lead to a breakdown of
%KK-parity and will allow, for example, single production of $n =
%1$ KK-excitations and their decay to SM particles. Earlier  we have
%examined, $B^1 (W_3^1) \rightarrow e^+e^-, \mu^+\mu^-$, decays
%after the production of the $B^1 (W_3^1)$ singly at the LHC \cite{ddrs}.

With positive values of BLKTs, we have studied electroweak interaction, in two alternative ways. In the first case we put equal strengths of fermion BLKTs at the two
fixed points and parametrised by $r_f$, while for electroweak gauge boson  we have considered unequal strengths of BLKTs $(r_V^a \neq r_V^b)$. Equal strengths of electroweak gauge boson BLKTs 
would preserve the $Z_2$-parity. In the other situation we have considered the fermion and electroweak gauge boson BLKTs are
present only at the $y = 0$ fixed point. These BLKTs modify the field equations and the boundary conditions of the solutions lead to the non-trivial KK-mass excitations and wave-functions of fermions and the electroweak gauge bosons in the $y$-direction in both cases. In this platform we have calculated KK-parity-non-conserving coupling between the $n = 1$ KK-excitation of the electroweak gauge bosons and pair of SM fermions ($n = 0$) in terms of $r_f, r^a_V, r^b_V ~{\rm and} ~1/R$ when BLKTs are present at both fixed points and $r_f, r_V, ~{\rm and} ~1/R$ for the other case. This driving coupling vanishes in the $\Delta R_V = 0$ limit in the first case and for $R_f = R_V$ in the second.

Finally we estimate the single production of $V^1$ at the LHC
and its subsequent decay to $\ell^{+}\ell^{-}$, both the production and decay are controlled by the KK-parity-non-conserving coupling.
We compare our results with the $\ell^{+}\ell^{-}$
resonance production signature at the  LHC running at 8 TeV $pp$
centre of momentum energy  \cite{atlas8T, cms8T}. The lack of observation of this signal with  20 $fb^{-1}$ accumulated
luminosity by ATLAS collaboration \cite{atlas8T} at the LHC already excludes a large part of the parameter space
(spanned by $r_f, r^a_V, r^b_V$ and $1/R$ in one case and $r_f,
r_V$ and $1/R$ in the other). Here we consider 
the $B^1 (W_3^1)$ is lighter than the corresponding fermion and the
bounds on the mass of the former are the same as that on the $\ell^{+}\ell^{-}$ resonance from the data. 

At the end, we also like to point out another important observation regarding the excluded parameter space of this model that the $\ell^{+}\ell^{-}$ resonance search disfavoured more parameter space in comparison to the $t\bar{t}$ resonance search which we performed in our previous article \cite{drs}. 

%The cross section limits from
%LHC put tight upper bounds on the mass of $n = 1$ fermion. In
%particular, while a range of a few hundred GeV  is still permitted
%for this mass in the first scenario, in the second the $n = 1$
%fermions and gauge bosons have to be quasi-degenerate. 

{\bf Acknowledgements} The author thanks Anindya Datta, Amitava Raychaudhuri and Ujjal Kumar Dey for many useful discussions. He is grateful to Debabrata Adak for carefully reading the manuscript and several illuminating suggestions. He is the recipient of a Senior Research Fellowship from the University Grants Commission.

\end{document}